\RequirePackage{fixltx2e}
\documentclass[a4paper]{isp}
\usepackage[american]{babel}
\usepackage{graphicx}
\usepackage{amsmath,amsthm,mathtools}
\usepackage{url}
\usepackage[capitalise,nameinlink]{cleveref}
\usepackage{xcolor}
\def\aap{A\&A\,  }

\def\aj{AJ  }

\def\mnras{MNRAS\,  }

\def\sun{\hbox{$\odot$}}
\def\twopar{TPLD}
\def\power{PLD}
\def\generalized{GLD}
\def\newgen{NGLD}
\def\weighted{NWL}
\def\truncated{DTL}
\def\sunmass{\mathcal{M}_{\sun}}

\begin{document}
\pdfgentounicode=1
\title
{
New probability distributions in astrophysics:
II. The   generalized and double truncated Lindley
}
\author{Lorenzo  Zaninetti}
\institute{
Physics Department,
 via P.Giuria 1, I-10125 Turin,Italy \\
 \email{zaninetti@ph.unito.it}
}

\maketitle

\begin {abstract}

The statistical parameters of five generalizations of
the Lindley distribution, such as the average, variance and 
moments, are reviewed.
A new double truncated Lindley distribution with three 
parameters is derived.
The new distributions are applied to model
the initial mass function for stars.
\end{abstract}

\section{Introduction}

The Lindley distribution, after \cite{Lindley1958,Lindley1965},
has one parameter.
In recent years the Lindley distribution 
has been the subject of many generalizations, we report
some of them among others:
one with two parameters  \cite{Shanker2013},
a two-parameter weighted one \cite{Ghitany2011},
the generalized Poisson--Lindley  \cite{Mahmoudi2010},
the extended Lindley  \cite{Bakouch2012} and 
a transmuted Lindley-geometric distribution \cite{Merovci2013}.
Several generalizations of the Lindley distribution
can be found in a recent review \cite{Tomy2018}.
The Lindley distribution  is 
useful in modeling biological data from grouped mortality studies
\cite{Ghitany2011,Shanker2014}
and the first application  to astrophysics of the 
Lindley distribution 
has been done for the initial  mass function (IMF) for stars
and the luminosity function for galaxies \cite{Zaninetti2019a}.
The IMF is routinely modeled by the lognormal distribution
and therefore the following question naturally arises.
Can a Lindley distribution or a generalization
be an alternative to the lognormal fit for  the IMF?
In order to answer the above  question 
Section  \ref{section_preliminaries} reviews the notion of 
statistical sample and Lindley distribution,
Section  \ref{section_generalizations} reviews 
five generalizations of the Lindley distribution,
Section  \ref{section_double} introduces 
the double Lindley distribution
and  
Section  \ref{section_applications}
fits  the six new Lindley distributions
to four samples for the mass of the stars. 

\section{Preliminaries}

\label{section_preliminaries}
We report some basic information on the adopted 
sample and on the original Lindley distribution 
with one parameter.

\subsection{The sample}

The experimental sample  consists of the data $x_i$ with
$i$  varying between  1 and  $n$;
the sample mean, $\bar{x}$,
is
\begin{equation}
\bar{x} =\frac{1}{n} \sum_{i=1}^{n} x_i
\quad ,
\label{xmsample}
\end{equation}
the unbiased sample variance, $s^2$, is
\begin{equation}
s^2 = \frac{1}{n-1}  \sum_{i=1}^{n} (x_i - \bar{x})^2
\quad ,
\label{variancesample}
\end{equation}
and the sample $r$th moment  about the origin, $\bar{x}_r$,
is
\begin{equation}
\bar{x}_r = \frac{1}{n} \sum_{i=1}^{n} (x_i)^r
\quad .
\label{rmoment}
\end{equation}

\subsection{The Lindley distribution with one parameter}

\label{lindleysec}
The {\em Lindley}
probability density function (PDF) with one parameter, $f(x)$,
is
\begin{equation}
f (x;c) =
\frac
{
{c}^{2}{{\rm e}^{-cx}} \left( x+1 \right)
}
{
1+c
}
\quad ; x>0 , c>0 
\end{equation}
where  $x>0$  and  $c>0$.

The  cumulative distribution function (CDF), $F(x)$,
 is
\begin{equation}
F (x;c) =
1- \left( 1+{\frac {cx}{1+c}} \right) {{\rm e}^{-cx}}
\quad ; x>0 , c>0 
\quad .
\end{equation}
At $x=0$, $f(0) = {\frac {{c}^{2}}{1+c}}$ and is not zero.

The average  value or mean, $\mu$,  is
\begin{equation}
\mu (c)=
\frac
{
2+c
}
{
c \left( 1+c \right)
}
\quad ,
\end{equation}
the variance, $\sigma^2$, is
\begin{equation}
\sigma^2(c)=
\frac
{
{c}^{2}+4\,c+2
}
{
{c}^{2} \left( 1+c \right) ^{2}
}
\quad .
\end{equation}
The $r$th moment about the origin for the Lindley distribution,
$\mu^{\prime}_r$, 
is
\begin{equation}
\mu^{\prime}_r =
\frac
{
{c}^{-r}\Gamma \left( r+2 \right) +{c}^{1-r}\Gamma \left( r+1 \right)
}
{
1+c
}
\quad ,
\end{equation}
where
\begin{equation}
\mathop{\Gamma\/}\nolimits\!\left(z\right)
=\int_{0}^{\infty}e^{{-t}}t^{{z-1}}dt
\quad ,
\end{equation}
is the gamma function, see \cite{NIST2010}.
The central moments, $\mu_r$, are
\begin{subequations}
\begin{align}
\mu_3& =
\frac
{
2\,{c}^{3}+12\,{c}^{2}+12\,c+4
}
{
{c}^{3} \left( 1+c \right) ^{3}
}  \\
\mu_4& =
\frac
{
9\,{c}^{4}+72\,{c}^{3}+132\,{c}^{2}+96\,c+24
}
{
{c}^{4} \left( 1+c \right) ^{4}
}
\end{align}
\end{subequations}
More details can be found in  \cite{Lindley1965}.

\section{Generalizations of the Lindley distribution}

\label{section_generalizations}
We review the statistics of  the
Lindley distribution  
with two parameters,
power,
generalized,
new  generalized and
new weighted.

\subsection{The Lindley distribution with two parameters}

The {\em Lindley}
PDF with two parameters \twopar{}  \cite{Shanker2013}  is
\begin{equation}
f (x;b,c) =
\frac
{
{c}^{2} \left( b+x \right) {{\rm e}^{-cx}}
}
{
bc+1
}
\quad ,
\end{equation}
where  $x>0$, $c>0$   and  $b\,c>-1$.
The CDF of the \twopar{} is
\begin{equation}
F(x;c,b)=
1-{\frac { \left( bc+cx+1 \right) {{\rm e}^{-cx}}}{bc+1}}
\quad .
\end{equation}
The average  value or mean of the \twopar{}  is
\begin{equation}
\mu (b,c)=
\frac
{
bc+2
}
{
c \left( bc+1 \right) 
}
\quad ,
\end{equation}
and 
the variance of the \twopar{}  is
\begin{equation}
\sigma^2(b,c)=
\frac
{
{b}^{2}{c}^{2}+4\,bc+2
}
{
{c}^{2} \left( bc+1 \right) ^{2}
}
\quad .
\end{equation}
The mode of the \twopar{} is  at
\begin{equation}
Mode={\frac {1-bc}{c}}
\quad,
\end{equation}
see eqn. (2.3) in \cite{Shanker2013}.
The $r$th moment about the origin for the \twopar{},
$\mu^{\prime}_r$,
is
\begin{equation}
\mu^{\prime}_r =
\frac
{
{c}^{1-r}b\Gamma \left( r+1 \right) +{c}^{-r}\Gamma \left( r+2
 \right) 
}
{
bc+1
}
\quad .
\end{equation}
The two parameters $b$ and  $c$ can be obtained by the following
match
\begin{subequations}
\begin{align}
\mu         & = \bar{x} 
\label{abmu} 
 \\
\sigma^2    & = s^2 \quad ,
\label{absigma2}
\end{align}
\end{subequations}
which means
\begin{equation}
\widehat{b}
=
\frac
{
- \left( {s}^{2}+{{\it \bar{x}}}^{2} \right)  \left( {\it \bar{x}}\,\sqrt {-2\,{
s}^{2}+2\,{{\it \bar{x}}}^{2}}-2\,{s}^{2} \right) 
}
{
\left( {\it \bar{x}}\,\sqrt {-2\,{s}^{2}+2\,{{\it \bar{x}}}^{2}}+{{\it \bar{x}}}^{2}
-{s}^{2} \right)  \left( 2\,{\it \bar{x}}+\sqrt {-2\,{s}^{2}+2\,{{\it \bar{x}}}^
{2}} \right) 
}
\quad ,
\end{equation}
and
\begin{equation}
\widehat{c}
=
\frac
{
2\,{\it \bar{x}}+\sqrt {-2\,{s}^{2}+2\,{{\it \bar{x}}}^{2}}
}
{
{s}^{2}+{{\it \bar{x}}}^{2}
}
\quad .
\end{equation}

\subsection{The power Lindley distribution }

The {\em power Lindley}
PDF with two parameters (\power) according to   
\cite{Shanker2013}  is
\begin{equation}
f (x;b,c) =
\frac
{
c{b}^{2} \left( 1+{x}^{c} \right) {x}^{c-1}{{\rm e}^{-b{x}^{c}}}
}
{
b+1
}
\quad ,
\end{equation}
where $b$, $c$ and $x>0$.
The CDF  of the \power{}  is
\begin{equation}
F(x;c,b)=
{\frac { \left( -b{x}^{c}-b-1 \right) {{\rm e}^{-b{x}^{c}}}+b+1}{b+1}}
\quad .
\end{equation}
The average  value or mean of the \power{}  is
\begin{equation}
\mu (b,c)=
\frac
{
\left( {b}^{-{c}^{-1}}c+{b}^{{\frac {c-1}{c}}}c+{b}^{-{c}^{-1}}
 \right) \Gamma \left( {\frac {c+1}{c}} \right) 
}
{
\left( b+1 \right) c
}
\quad ,
\end{equation}
and 
the variance of the \power{}  is
\begin{equation}
\sigma^2(b,c)=
\frac
{
NA
}
{
DA
}
\quad ,
\end{equation}
where 
\begin{eqnarray}
NA=
-{b}^{-2\,{c}^{-1}} \left( \Gamma \left( {\frac {c+1}{c}} \right) 
 \right) ^{2}{c}^{2}+{b}^{-2\,{c}^{-1}}\Gamma \left( {\frac {c+2}{c}}
 \right) b{c}^{2}-{b}^{{\frac {2\,c-2}{c}}} \left( \Gamma \left( {
\frac {c+1}{c}} \right)  \right) ^{2}{c}^{2}
\nonumber \\
-2\, \left( \Gamma \left( 
{\frac {c+1}{c}} \right)  \right) ^{2}{b}^{{\frac {-2+c}{c}}}{c}^{2}+{
b}^{{\frac {-2+c}{c}}}\Gamma \left( {\frac {c+2}{c}} \right) b{c}^{2}-
2\,{b}^{-2\,{c}^{-1}} \left( \Gamma \left( {\frac {c+1}{c}} \right) 
 \right) ^{2}c
\nonumber \\
+2\,{b}^{-2\,{c}^{-1}}\Gamma \left( {\frac {c+2}{c}}
 \right) bc+{b}^{-2\,{c}^{-1}}\Gamma \left( {\frac {c+2}{c}} \right) {
c}^{2}-2\, \left( \Gamma \left( {\frac {c+1}{c}} \right)  \right) ^{2}
{b}^{{\frac {-2+c}{c}}}c
\nonumber \\
+{b}^{{\frac {-2+c}{c}}}\Gamma \left( {\frac {
c+2}{c}} \right) {c}^{2}-{b}^{-2\,{c}^{-1}} \left( \Gamma \left( {
\frac {c+1}{c}} \right)  \right) ^{2}+2\,{b}^{-2\,{c}^{-1}}\Gamma
 \left( {\frac {c+2}{c}} \right) c
\quad ,
\end{eqnarray}
and
\begin{equation}
DA=
\left( b+1 \right) ^{2}{c}^{2}
\quad .
\end{equation}
The mode of the \power{} is  at
\begin{equation}
Mode=
\frac
{
-cb+\sqrt {1+ \left( {b}^{2}+4 \right) {c}^{2}+ \left( -2\,b-4
 \right) c}+2\,c-1
}
{
2\,cb
}
\quad.
\end{equation}
The $r$th moment about the origin for the \power{}  is
\begin{equation}
\mu^{\prime}_r =
\frac
{
{b}^{{\frac {-r+c}{c}}}\Gamma \left( {\frac {r+c}{c}} \right) +{b}^{-{
\frac {r}{c}}}\Gamma \left( {\frac {r+2\,c}{c}} \right) 
}
{
b+1
}
\quad .
\end{equation}
The two parameters $b$ and $c$ of the \power{} can be found  
by numerically solving the nonlinear system  given by eqs (\ref{abmu}) and
(\ref{absigma2}).

\subsection{The generalized  Lindley distribution }

The {\em generalized  Lindley}
PDF with three parameters (\generalized) according to   
\cite{Zakerzadeh2009}  is
\begin{equation}
f (x;a,b,c) =
\frac
{
{b}^{2} \left( bx \right) ^{a-1} \left( cx+a \right) {{\rm e}^{-bx}}
}
{
\left( c+b \right) \Gamma \left( a+1 \right) 
}
\quad ,
\end{equation}
where $a$, $b$, $c$ and $x>0$.
The CDF  of the \generalized{}  is
\begin{equation}
F(x;a,c,b)=
{\frac {{{\rm e}^{-1/2\,bx}} \left( {x}^{a/2} \left( c{b}^{a/2}+{b}^{a
/2+1} \right) {{\sl M}_{a/2,\,a/2+1/2}\left(bx\right)}+{b}^{a+1}{x}^{a
}{{\rm e}^{-1/2\,bx}} \left( a+1 \right)  \right) }{ \left( c+b
 \right) \Gamma \left( a+2 \right) }}
\quad ,
\label{dfgeneralized}
\end{equation}
where ${{\sl M}_{\mu,\,\nu}\left(z\right)}$ is 
the Whittaker $M$ function, see \cite{NIST2010}.
The average  value or mean of the \generalized{}  is
\begin{equation}
\mu (a,b,c)=
\frac
{
ab+ac+c
}
{
b \left( c+b \right)
}
\quad ,
\end{equation}
and 
the variance of the \generalized{}  is
\begin{equation}
\sigma^2(a,b,c)=
\frac
{
a{b}^{2}+2\,cba+{c}^{2}a+2\,cb+{c}^{2}
}
{
{b}^{2} \left( c+b \right) ^{2}
}
\quad .
\end{equation}

\begin{figure}
 \centering
\includegraphics[width=6cm]{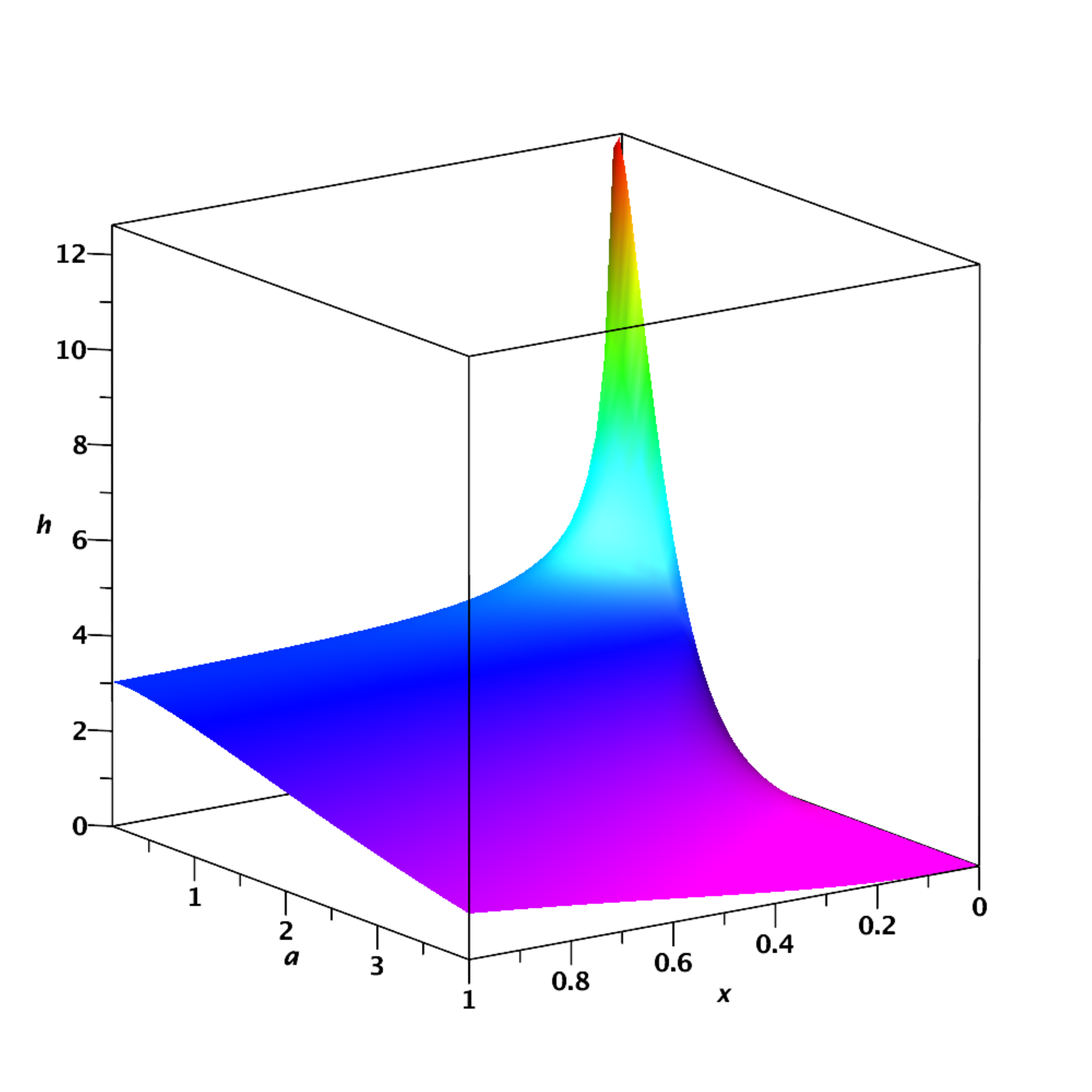}
\caption 
{
Plot of the three-dimensional surface
of the hazard rate function when $b$=3
and $c$=0.5.
}
 \label{hazard} 
 \end{figure}

The hazard rate function, $h(x;a,b,c)$, of the \generalized{} 
is
\begin{eqnarray}
h(x;a,b,c) =
\nonumber \\
\frac
{
-{b}^{a+1}{x}^{a-1} \left( cx+a \right) {{\rm e}^{-bx}} \left( a+1
 \right) 
}
{
{{\rm e}^{-1/2\,bx}}{x}^{a/2} \left( c{b}^{a/2}+{b}^{a/2+1} \right) {
{\sl M}_{a/2,\,a/2+1/2}\left(bx\right)}+{x}^{a}{b}^{a+1} \left( a+1
 \right) {{\rm e}^{-bx}}- \left( c+b \right) \Gamma \left( a+2
 \right) 
}
\quad ,
\label{hgeneralized}
\end{eqnarray}
and  
Figure \ref{hazard} reports an example.
Here the CDF, equation (\ref{dfgeneralized}),
and the hazard rate function, equation (\ref{hgeneralized}),
are reported in closed form in contrast to
what was asserted by \cite{Zakerzadeh2009}.
The mode of the \generalized{} is  at
\begin{equation}
Mode=
\frac
{
-ab+ac+\sqrt {{a}^{2}{b}^{2}+2\,{a}^{2}bc+{a}^{2}{c}^{2}-4\,abc}
}
{
2\,bc
}
\quad .
\end{equation}
The $r$th moment about the origin for the \generalized{}  is
\begin{equation}
\mu^{\prime}_r =
\frac
{
\Gamma \left( r+a \right)  \left( {b}^{-r}ca+{b}^{-r}cr+{b}^{-r+1}a
 \right) 
}
{
\left( c+b \right) \Gamma \left( a+1 \right)
}
\quad ,
\end{equation}
and  in particular the third moment is
\begin{equation}
\mu^{\prime}_3 =
\frac
{
\Gamma \left( 3+a \right)  \left( ab+ac+3\,c \right) 
}
{
\left( c+b \right) \Gamma \left( a+1 \right) {b}^{3}
}
\quad .
\end{equation}
The three  parameters $a$, $b$ and  $c$ of the 
\generalized{}
can be obtained by 
numerically solving the following three non-linear equations
\begin{subequations}
\begin{align}
\mu            & = \bar{x} 
\label{eqn1}
 \\
\sigma^2       & = s^2
\label{eqn2} 
\\
\mu^{\prime}_3 & = \bar{x}_3
\label{eqn3}
\quad .
\end{align}
\end{subequations}

\subsection{The new generalized  Lindley distribution }

The {\em new generalized  Lindley}
PDF with three parameters (\newgen) according to   
\cite{Ibrahim2013}  is
\begin{equation}
f (x;a,b,c) =
\frac
{
\left( {c}^{a+1}{x}^{a-1}\Gamma \left( b \right) +{c}^{b}{x}^{b-1}
\Gamma \left( a \right)  \right) {{\rm e}^{-cx}}
}
{
 \left( 1+c \right) \Gamma \left( a \right) \Gamma \left( b \right) 
}
\quad ,
\end{equation}
where $a$, $b$, $c$ and $x>0$.
The CDF  of the \newgen{}  is
\begin{equation}
F(x;a,c,b)=
\frac
{NB
}
{
\left( 1+c \right) \Gamma \left( b+2 \right) \Gamma \left( a+2
 \right) 
}
\quad ,
\end{equation}
where 
\begin{eqnarray}
NB=\Gamma \left( b+2 \right) {x}^{a}{c}^{a+1}{{\rm e}^{-cx}}a+\Gamma
 \left( a+2 \right) {x}^{b}{c}^{b}{{\rm e}^{-cx}}b-\Gamma \left( b+2
 \right) c\Gamma \left( a+1,cx \right) a+
\nonumber \\
\Gamma \left( b+2 \right) {x}
^{a}{c}^{a+1}{{\rm e}^{-cx}}+\Gamma \left( a+2 \right) {x}^{b}{c}^{b}{
{\rm e}^{-cx}}-\Gamma \left( b+2 \right) c\Gamma \left( a+1,cx
 \right) +\Gamma \left( b+2 \right) \Gamma \left( a+2 \right) c-
\nonumber \\
\Gamma
 \left( a+2 \right) \Gamma \left( b+1,cx \right) b+\Gamma \left( b+2
 \right) \Gamma \left( a+2 \right) -\Gamma \left( a+2 \right) \Gamma
 \left( b+1,cx \right) 
\end{eqnarray}
where $\Gamma(a, z)$ is the incomplete Gamma function,
defined by
\begin{equation}
\mathop{\Gamma\/}\nolimits\!\left(a,z\right)
=\int_{z}^{\infty}t^{a-1}e^{-t}dt
\quad ,
\end{equation}
see  \cite{NIST2010}.
The average  value  of the \newgen{}  is
\begin{equation}
\mu (a,b,c)=
\frac
{
ac+b
}
{
c \left( 1+c \right)
}
\quad ,
\end{equation}
and the variance of the \newgen{}  is
\begin{equation}
\sigma^2(a,b,c)=
\frac
{
{a}^{2}c-2\,abc+a{c}^{2}+{b}^{2}c+ac+bc+b
}
{
{c}^{2} \left( 1+c \right) ^{2}
}
\quad .
\end{equation}

The $r$th moment about the origin for the \newgen{}  is
\begin{equation}
\mu^{\prime}_r =
\frac
{
{c}^{-r+1}\Gamma \left( r+a \right) \Gamma \left( b \right) +{c}^{-r}
\Gamma \left( r+b \right) \Gamma \left( a \right) 
}
{
\left( 1+c \right) \Gamma \left( a \right) \Gamma \left( b \right) 
}
\quad ,
\end{equation}
and   the third moment is
\begin{equation}
\mu^{\prime}_3 =
\frac
{
\Gamma \left( 3+a \right) \Gamma \left( b \right) c+\Gamma \left( 3+b
 \right) \Gamma \left( a \right) 
}
{
{c}^{3} \left( 1+c \right) \Gamma \left( a \right) \Gamma \left( b
 \right) 
}
\quad .
\end{equation}
The three  parameters $a$, $b$ and  $c$ of the 
\newgen{}
are  obtained by 
numerically solving the  three non-linear equations
(\ref{eqn1}), (\ref{eqn2}) and (\ref{eqn1}).

\subsection{The new weighted Lindley  distribution }

The {\em new weighted Lindley}
PDF with two parameters (\weighted) according to   
\cite{Asgharzadeh2016}  is
\begin{equation}
f (x;b,c) =
\frac
{
-{c}^{2} \left( 1+b \right) ^{2} \left( 1+x \right)  \left( -1+{
{\rm e}^{-cbx}} \right) {{\rm e}^{-cx}}
}
{
b \left( cb+b+c+2 \right) 
}
\quad ,
\end{equation}
where  $b$, $c$ and $x>0$.
The CDF  of the \weighted{}  is
\begin{equation}
F(x;c,b)=
\frac
{
NC
}
{
b \left( cb+b+c+2 \right)
}
\quad ,
\end{equation}
where 
\begin{eqnarray}
NC=
-{{\rm e}^{-cx}}{b}^{2}cx+{{\rm e}^{-c \left( 1+b \right) x}}bcx-{
{\rm e}^{-cx}}{b}^{2}c-2\,{{\rm e}^{-cx}}bcx+{{\rm e}^{-c \left( 1+b
 \right) x}}bc
\nonumber \\
+{{\rm e}^{-c \left( 1+b \right) x}}cx
-{{\rm e}^{-cx}}{b
}^{2}-2\,{{\rm e}^{-cx}}bc-{{\rm e}^{-cx}}cx+{b}^{2}c+{{\rm e}^{-c
 \left( 1+b \right) x}}c-2\,{{\rm e}^{-cx}}b-{{\rm e}^{-cx}}c+{b}^{2}+
cb
\nonumber \\
+{{\rm e}^{-c \left( 1+b \right) x}}-{{\rm e}^{-cx}}+2\,b
\quad .
\end{eqnarray}
The average  value  of the \weighted{}  is
\begin{equation}
\mu (a,b,c)=
\frac
{
{b}^{2}c+2\,{b}^{2}+3\,cb+6\,b+2\,c+6
}
{
\left( cb+b+c+2 \right) c \left( 1+b \right)
}
\quad ,
\end{equation}
and the variance of the \weighted{}  is
\begin{equation}
\sigma^2(a,b,c)=
\frac
{
ND
}
{
{c}^{2} \left( bc+b+c+2 \right) ^{2} \left( 1+b \right) ^{2}
}
\quad ,
\end{equation}
where 
\begin{eqnarray}
ND =
{b}^{4}{c}^{2}+4\,{b}^{4}c+4\,{b}^{3}{c}^{2}+2\,{b}^{4}+18\,{b}^{3}c+7
\,{b}^{2}{c}^{2}+12\,{b}^{3}+32\,{b}^{2}c+6\,b{c}^{2}+24\,{b}^{2}
\nonumber \\
+30\,
bc+2\,{c}^{2}+24\,b+12\,c+12
\quad .
\end{eqnarray}
The $r$th moment about the origin for the \weighted{}  is
\begin{equation}
\mu^{\prime}_r =
\frac
{
NE
}
{
b \left( bc+b+c+2 \right) 
}
\quad ,
\end{equation}
where
\begin{eqnarray}
NE=
-  \Big ( {c}^{1-r}{b}^{1-r}   ( {\frac {1+b}{b}}   ) ^{-r}+{b}^
{-r}   ( {\frac {1+b}{b}}   ) ^{-r}{c}^{-r}r-{c}^{-r}{b}^{2}r+{
c}^{1-r}{b}^{-r}   ( {\frac {1+b}{b}}   ) ^{-r}
-{c}^{1-r}{b}^{2
}
\nonumber \\
+{b}^{-r}   ( {\frac {1+b}{b}}   ) ^{-r}{c}^{-r}-{c}^{-r}{b}^{
2}-2\,{c}^{-r}br-2\,{c}^{1-r}b-2\,{c}^{-r}b-{c}^{-r}r-{c}^{1-r}-{c}^{-
r}  \Big  ) \Gamma   ( 1+r   ) 
\quad .
\end{eqnarray}
The two parameters $b$ and $c$ of the \weighted{} can be found  
by numerically solving the nonlinear system  given by eqs (\ref{abmu}) and
(\ref{absigma2}).

\section{The  double truncated Lindley distribution}

\label{section_double}
Let $X$ be a random variable
defined in
$[x_l,x_u]$;
the {\em double truncated } (\truncated)
version of the Lindley PDF with one parameter, 
$f_t(x;c,x_{l},x_{u})$,
is
\begin{equation}
f_t(x;c,x_{l},x_{u})=
\frac
{
{c}^{2}{{\rm e}^{-cx}} \left( x+1 \right) 
}
{
{{\rm e}^{-c{\it x_l}}}c{\it x_l}-{{\rm e}^{-c{\it x_u}}}c{\it x_u}+{
{\rm e}^{-c{\it x_l}}}c-{{\rm e}^{-c{\it x_u}}}c+{{\rm e}^{-c{\it x_l}}}-
{{\rm e}^{-c{\it x_u}}}
}
\quad ,
\label{pdflindleytruncated}
\end{equation}
where the effect of the double truncation increases 
the parameters from one to
three, see \cite{Singh2014}.
The double truncated Lindley distribution with scale,
which has four parameters, was introduced  in \cite{Zaninetti2019a}.

Its CDF, $F_t(x;b,c,x_{l},x_{u})$, is  
\begin{equation}
F_t(x;b,c,x_{l},x_{u})
=
\frac
{
NF
}
{
 \left(  \left( -1+ \left( -{\it x_u}-1 \right) c \right) {{\rm e}^{c{
\it x_l}}}+ \left( 1+ \left( {\it x_l}+1 \right) c \right) {{\rm e}^{c{
\it x_u}}} \right) ^{2}
}
\quad  ,
\end{equation}
where 
\begin{eqnarray}
NF = 
-{{\rm e}^{c   ( {\it x_l}+{\it x_u}   ) }} \Big  ( -   ( 1+
   ( {\it x_l}+1   ) c   ) ^{2}{{\rm e}^{-c   ( {\it x_l}-
{\it x_u}   ) }}-   ( 1+   ( x+1   ) c   )    ( 1+
   ( {\it x_u}+1   ) c   ) {{\rm e}^{c   ( -x+{\it x_l}
   ) }}
\nonumber  \\
+  \Big (    ( 1+   ( x+1   ) c   ) {{\rm e}^{
c   ( -x+{\it x_u}   ) }}+1+   ( {\it x_u}+1   ) c
  \Big )    ( 1+   ( {\it x_l}+1   ) c   )   \Big ) 
\quad .
\end{eqnarray}
The average value, $\mu_t(c,x_{l},x_{u})$,
is   
\begin{equation}
\mu_t(c,x_{l},x_{u})
=
\frac
{
\left( 2+ \left( {{\it x_u}}^{2}+{\it x_u} \right) {c}^{2}+ \left( 2\,{
\it x_u}+1 \right) c \right) {{\rm e}^{c{\it x_l}}}-{{\rm e}^{c{\it x_u}}
} \left( 2+ \left( {{\it x_l}}^{2}+{\it x_l} \right) {c}^{2}+ \left( 2\,
{\it x_l}+1 \right) c \right)
}
{
-c \left(  \left( -1+ \left( -{\it x_u}-1 \right) c \right) {{\rm e}^{c
{\it x_l}}}+{{\rm e}^{c{\it x_u}}} \left( 1+ \left( {\it x_l}+1 \right) c
 \right)  \right) 
}
\quad  .
\end{equation}
The $r$th moment about the origin for the \truncated{},
$\mu^{\prime}_r(c,x_l,x_u)$, 
is
\begin{equation}
\mu^{\prime}_r(c,x_l,x_u)=
\frac
{
NG
}
{
\left(  \left( 1+ \left( {\it x_l}+1 \right) c \right) {{\rm e}^{-c{
\it x_l}}}- \left( 1+ \left( {\it x_u}+1 \right) c \right) {{\rm e}^{-c{
\it x_u}}} \right)  \left( r+1 \right) 
}
\quad ,
\end{equation}
where
\begin{eqnarray}
NG =
-{{\it x_l}}^{r/2}{{\rm e}^{-1/2 c{\it x_l}}} \left( {c}^{1-r/2}+{c}^{-
r/2} \left( r+1 \right)  \right) {{\sl M}_{r/2, r/2+1/2}\left(c{\it 
x_l}\right)}
\nonumber \\
+ \left( {c}^{1-r/2}+{c}^{-r/2} \left( r+1 \right) 
 \right) {{\rm e}^{-1/2 c{\it x_u}}}{{\it x_u}}^{r/2}{{\sl M}_{r/2, r/
2+1/2}\left(c{\it x_u}\right)}
\nonumber \\
+c \left( r+1 \right)  \left( {{\rm e}^{-
c{\it x_l}}}{{\it x_l}}^{r+1}-{{\rm e}^{-c{\it x_u}}}{{\it x_u}}^{r+1}
 \right) 
\quad  .
\end{eqnarray}
The three parameters which characterize the \truncated{} 
can be found in the following way.
Consider the   sample of stellar masses  
${\mathcal X}=x_1, x_2, \dots , x_n$ and let
$x_{(1)} \geq x_{(2)} \geq \dots \geq x_{(n)}$ denote
their order statistics, so that
$x_{(1)}=\max(x_1, x_2, \dots, x_n)$, $x_{(n)}=\min(x_1, x_2, \dots, x_n)$.
The first two parameters $x_l$ and $x_u$
are
\begin{equation}
{x_l}=x_{(n)}, \qquad { x_u}=x_{(1)}
\quad  .
\label{eq:firstpar}
\end{equation}
The third parameter $c$ can be found by solving the 
following non-linear equation 
\begin{equation}
\mu_t(c,x_{l},x_{u}) = \bar{x} 
\quad .
\end{equation}

\section{Application to the IMF}

\label{section_applications}
We report the adopted statistics for
four samples of stars which will be subject of fit,
 with 
the lognormal,
the Lindley generalizations
and
the double truncated Lindley.

\subsection{The involved  statistics}

The merit function $\chi^2$
is computed
according to the formula
\begin{equation}
\chi^2 = \sum_{i=1}^n \frac { (T_i - O_i)^2} {T_i},
\label{chisquare}
\end {equation}
where $n  $   is the number of bins,
      $T_i$   is the theoretical value,
and   $O_i$   is the experimental value represented
by the frequencies.
The theoretical  frequency distribution is given by
\begin{equation}
 T_i  = N {\Delta x_i } p(x) \quad,
\label{frequenciesteo}
\end{equation}
where $N$ is the number of elements of the sample,
      $\Delta x_i $ is the magnitude of the size interval,
and   $p(x)$ is the PDF  under examination.

A reduced  merit function $\chi_{red}^2$
is  evaluated  by
\begin{equation}
\chi_{red}^2 = \chi^2/NF
\quad,
\label{chisquarereduced}
\end{equation}
where $NF=n-k$ is the number of degrees  of freedom,
$n$ is the number of bins,
and $k$ is the number of parameters.
The goodness  of the fit can be expressed by
the probability $Q$, see  equation 15.2.12  in \cite{press},
which involves the degrees of freedom
and $\chi^2$.
According to  \cite{press} p.~658, the
fit `may be acceptable' if  $Q>0.001$. 

The Akaike information criterion
(AIC), see \cite{Akaike1974},
is defined by
\begin{equation}
AIC  = 2k - 2  ln(L)
\quad,
\end {equation}
where $L$ is
the likelihood  function  and $k$  the number of  free parameters
in the model.
We assume  a Gaussian distribution for  the errors
and  the likelihood  function
can be derived  from the $\chi^2$ statistic
$L \propto \exp (- \frac{\chi^2}{2} ) $
where  $\chi^2$ has been computed by
eqn.~(\ref{chisquare}),
see~\cite{Liddle2004}, \cite{Godlowski2005}.
Now the AIC becomes
\begin{equation}
AIC  = 2k + \chi^2
\quad.
\label{AIC}
\end {equation}

The Kolmogorov--Smirnov test (K--S),
see \cite{Kolmogoroff1941,Smirnov1948,Massey1951},
does not  require binning the data.
The K--S test,
as implemented by the FORTRAN subroutine KSONE in \cite{press},
finds
the maximum  distance, $D$, between the theoretical
and the astronomical  CDF
as well the  significance  level  $P_{KS}$,
see formulas  14.3.5 and 14.3.9  in \cite{press};
if  $ P_{KS} \geq 0.1 $,
the goodness of the fit is believable.

\subsection{The selected sample of stars}

\label{applications}

The {\it first}  test  is performed
on  NGC 2362  where
the  271 stars
have a range
 $1.47  {M}_{\sun}~\geq~ {M} \geq  0.11  {M}_{\sun}$,
see  \cite{Irwin2008} and CDS catalog J/MNRAS/384/675/table1.

The {\it second}   test  is performed
on   the low-mass IMF
in the young cluster NGC 6611,
see  \cite{Oliveira2009} and CDS catalog J/MNRAS/392/1034.
This massive cluster has an age of 2--3 Myr and contains
masses from 
 $1.5  {M}_{\sun}~\geq~ {M} \geq  0.02  {M}_{\sun}$.
Therefore the brown dwarfs (BD)  region,
$\approx \, 0.2\,\sunmass$ is covered.

The {\it third }  test  is performed
on    $\gamma$ Velorum  cluster  where
the  237 stars
have a range
 $1.31  {M}_{\sun}~\geq~ {M} \geq  0.15  {M}_{\sun}$,
see  \cite{Prisinzano2016} and CDS catalog J/A+A/589/A70/table5.

The {\it fourth }  test  is performed
on     young cluster Berkeley 59  where
the  420 stars
have a range
 $2.24   {M}_{\sun}~\geq~ {M} \geq  0.15  {M}_{\sun}$,
see  \cite{Panwar2018} and CDS catalog J/AJ/155/44/table3.

\subsection{The lognormal distribution}

Let $X$ be a random variable
defined in
$[0, \infty]$;
the {\em lognormal}
PDF, 
following \cite{evans}
or formula (14.2)  in
\cite{univariate1}, is
\begin{equation}
PDF (x;m,\sigma) = \frac
{
{{\rm e}^{-\,{\frac {1}{{2\,\sigma}^{2}} \left( \ln  \left( {
\frac {x}{m}} \right )  \right ) ^{2}}}}
}
{
x\sigma\,\sqrt {2\,\pi}
}
\quad,
\label{pdflognormal}
\end{equation}
where $m$ is the median and $\sigma$ the shape parameter.
The CDF  is
\begin{equation}
CDF (x;m,\sigma) =
\frac{1}{2}+\frac{1}{2}\,{\rm erf} \left(\frac{1}{2}\,{\frac {\sqrt {2} \left( -\ln  \left( m
 \right ) +\ln  \left( x \right )  \right ) }{\sigma}}\right )
\quad ,
\end{equation}
where erf$(x)$ is the error function, defined as
\begin{equation}
\mathop{\mathrm{erf}\/}\nolimits
(x)=\frac{2}{\sqrt{\pi}}\int_{0}^{x}e^{-t^{2}}dt
\quad ,
\end{equation}
see \cite{NIST2010}.
The average  value or mean, $E(X)$,  is
\begin{equation}
E (X;m,\sigma) = m{{\rm e}^{\frac{1}{2}\,{\sigma}^{2}}}
\quad ,
\label{xmlognormal}
\end{equation}
the variance, $Var(X)$, is
\begin{equation}
Var=
{{\rm e}^{{\sigma}^{2}}} \left({{\rm e}^{{\sigma}^{2}}}-1 \right ) {m
}^{2}
\quad,
\label{varlognormal}
\end{equation}
the second moment about the origin, $E^2(X)$, is
\begin{equation}
E (X^2;m,\sigma) = {m}^{2}{{\rm e}^{2\,{\sigma}^{2}}}
\quad .
\label{momento2lognormal}
\end{equation}
The  statistics for the lognormal distribution 
for these four astronomical samples of stars 
are reported in Table \ref{chi2valueslognormal}.

\begin{table}[ht!]
\caption
{
Numerical values of
$\chi_{red}^2$, AIC, 
probability $Q$,
$D$, 
the maximum distance between theoretical and observed CDF,
and  $P_{KS}$, 
significance level,   in the K--S test of the lognormal distribution,
see equation~(\ref{pdflognormal}),
for  different  mass distributions.
The  number of  linear   bins, $n$, is 20.
}
\label{chi2valueslognormal}
\begin{center}
\begin{tabular}{|c|c|c|c|c|c|c|}
\hline
Cluster     &   parameters                           &  AIC   & $\chi_{red}^2$ & $Q$               &  $D$        &   $P_{KS}$  \\
\hline
NGC~2362    &  $\sigma$= 0.5,   $\mu_{LN} =-0.55 $   &  37.6  & 1.86           & 0.014             & 0.073     & 0.105     \\
NGC~6611    &  $\sigma$= 1.03 , $\mu_{LN} = -1.26$   &  71.2  & 3.73           & $1.31\,10^{-7}$   & 0.093     &  0.049 \\
$\gamma$~Velorum   &  $\sigma$= 0.5 ,  $\mu_{LN} = -1.08$   &  55.1  & 2.84           & $5.08\,10^{-5}$   & 0.092     &  0.033 \\
Berkeley 59 &  $\sigma$= 0.49 , $\mu_{LN} = -0.92$   &  54.9  & 2.82           & $5.49\,10^{-5}$   &  0.11     &  $6.46\,10^{-5}$ \\
\hline
\end{tabular}
\end{center}
\end{table}

\subsection{The generalizations of the Lindley distribution}

\begin{figure*}
\begin{center}
\includegraphics[width=6cm,angle=-90]{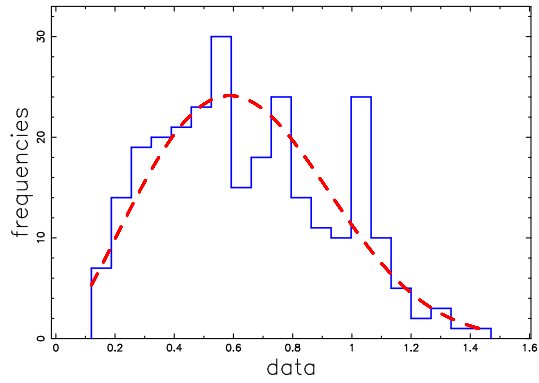}
\end{center}
\caption
{
Empirical PDF   of  mass distribution
for   NGC 2362 cluster data (273 stars + BDs)
when the number of bins, $n$, is 20
(steps with blue full line)
with a superposition of the \power{}  (red dashed line).
Theoretical parameters as in Table \ref{chi2lindleypower}.
}
\label{lindley_power_ngc2362}
\end{figure*}

The  statistics for the Lindley distribution 
and its  generalizations are reported in the following tables:
Table \ref{chi2lindleynorm} for  the Lindley distribution with one
parameter,
\begin{table}[ht!]
\caption
{
Numerical values of
$\chi_{red}^2$, AIC, 
probability $Q$,
$D$, 
the maximum distance between theoretical and observed CDF,
and  $P_{KS}$, 
significance level,   in the K--S test of the 
Lindley distribution with one parameter 
for  different  mass distributions.
The  number of  linear   bins, $n$, is 20.
}
\label{chi2lindleynorm}
\begin{center}
\begin{tabular}{|c|c|c|c|c|c|c|}
\hline
Cluster     &parameters &  AIC    & $\chi_{red}^2$ & $Q$               &  $D$       &   $P_{KS}$      \\
\hline
NGC~2362    &$c= 2.05$   &   95.57  &  5.03          & 3.36\,$10^{-12}$  & 0.248    & 2.93\,$10^{-15}$\\
NGC~6611    &$c= 2.94$   &   38.35  &  2.01          &  0.0053           & 0.077    & 0.161 \\
$\gamma$~Velorum    &$c= 3.18$   &   90.59  &  4.66          & $5.86\,10^{-11}$  & 0.322    & $3.23\,10^{-22}$ \\
Berkeley~59 &$c= 2.76$   &   149.6  &  7.76          & $6.35\,10^{-22}$  
& 0.323     & $5.24\,10^{-39}$ \\
\hline
\end{tabular}
\end{center}
\end{table}
Table \ref {chi2lindleytwopar} for the \twopar{}, 
\begin{table}[ht!]
\caption
{
Numerical values of
$\chi_{red}^2$, AIC, 
probability $Q$,
$D$, 
the maximum distance between theoretical and observed CDF,
and  $P_{KS}$, 
significance level,   in the K--S test of the 
\twopar{} distribution  with  two parameters 
for  different  mass distributions.
The  number of  linear   bins, $n$, is 20.
}
\label{chi2lindleytwopar}
\begin{center}
\begin{tabular}{|c|c|c|c|c|c|c|}
\hline
Cluster     &parameters &  AIC    & $\chi_{red}^2$ & $Q$               &  $D$       &   $P_{KS}$      \\
\hline
NGC~2362    &$b=-0.099, c= 4.2$ &  72.94  &  3.83    & 6.8\,$10^{-8}$    & 0.129    & 1.76\,$10^{-4}$\\
NGC~6611    &$b=0.043 , c= 4.32$&  59.11  &  3.06    & 1.23\,$10^{-5}$   & 0.098    & 0.033 \\
$\gamma$~Velorum    &$b=-0.035, c= 5.81$&  67.74  &  3.54    & $5\,10^{-7}$      & 0.14    & $8\,10^{-5}$ \\
Berkeley~59 &$b=-0.032, c= 4.75$&  81.47  &  4.3     & $2.35\,10^{-9}$   & 0.167     & $8.62\,10^{-11}$ \\
\hline
\end{tabular}
\end{center}
\end{table}
Table \ref {chi2lindleypower} for the \power{}, 
\begin{table}[ht!]
\caption
{
Numerical values of
$\chi_{red}^2$, AIC, 
probability $Q$,
$D$, 
the maximum distance between theoretical and observed CDF,
and  $P_{KS}$, 
significance level,   in the K--S test of the 
\power{} distribution  with  two parameters 
for  different  mass distributions.
The  number of  linear   bins, $n$, is 20.
}
\label{chi2lindleypower}
\begin{center}
\begin{tabular}{|c|c|c|c|c|c|c|}
\hline
Cluster     &parameters &  AIC    & $\chi_{red}^2$ & $Q$               &  $D$       &   $P_{KS}$      \\
\hline
NGC~2362    &$b=2.66, c= 2.28$  &   28.87  &  1.38   & 0.128             &
0.053    &   0.39 \\
NGC~6611    &$b=3.33 , c= 1.27$ &  53.53  &  2.75    & 8.88\,$10^{-5}$   &
0.087    & 0.08\\
$\gamma$~Velorum    &$b=4.64, c= 1.64$&  106.2  &   5.67    & $8.59 \,10^{-14}$      & 0.16    & $2\,10^{-6}$ \\
Berkeley~59 &$b=3.48, c= 1.54$  &  117.1  &  6.28     & $8\,10^{-16}$   
& 0.187     & $2.37\,10^{-13}$ \\
\hline
\end{tabular}
\end{center}
\end{table}
Table \ref {chi2lindleygeneralized} for the \generalized{}, 
\begin{table}[ht!]
\caption
{
Numerical values of
$\chi_{red}^2$, AIC, 
probability $Q$,
$D$, 
the maximum distance between theoretical and observed CDF,
and  $P_{KS}$, 
significance level,   in the K--S test of the 
\generalized{} distribution  with  three parameters 
for  different  mass distributions.
The  number of  linear   bins, $n$, is 20.
}
\label{chi2lindleygeneralized}
\begin{center}
\begin{tabular}{|c|c|c|c|c|c|c|}
\hline
Cluster     &parameters                  &  AIC    & $\chi_{red}^2$ & $Q$               &  $D$       &   $P_{KS}$      \\
\hline
NGC~2362    &$a=4.80, b=8.38, c= 12.01$  &   37.63  &  1.86   &0.016 &
0.064    &  0.2 \\
NGC~6611    &$a=1.4, b=4.8, c=8$ &  64.34  &  3.43    & 1.96\,$10^{-6}$   &
0.105   &  0.017\\
$\gamma$~Velorum    &$a=2.53, b=6.5, c= 0.00046$&   83.08  &  4.53    & $1.25
\,10^{-9}$      & 0.15    & $2.8\,10^{-5}$ \\
Berkeley~59 &$a=2.2, b=5.09, c= 1$  &  100.6  &  5.56     & $8.6\,10^{-13}$   
& 0.179    & $2.93\,10^{-12}$ \\
\hline
\end{tabular}
\end{center}
\end{table}
Table \ref {chi2lindleynewgeneralized} for the \newgen{} and  
\begin{table}[ht!]
\caption
{
Numerical values of
$\chi_{red}^2$, AIC, 
probability $Q$,
$D$, 
the maximum distance between theoretical and observed CDF,
and  $P_{KS}$, 
significance level,   in the K--S test of the 
\newgen{} distribution  with  three parameters 
for  different  mass distributions.
The  number of  linear   bins, $n$, is 20.
}
\label{chi2lindleynewgeneralized}
\begin{center}
\begin{tabular}{|c|c|c|c|c|c|c|}
\hline
Cluster     &parameters                  &  AIC    & $\chi_{red}^2$ & $Q$               &  $D$       &   $P_{KS}$      \\
\hline
NGC~2362    &$a=7.34, b=1.57, c= 10.61$  &  48.64  &  2.5   &5.4$10^{-4}$ &
0.075       &  0.086 \\
NGC~6611    &$a=3.14, b=-0.36, c= 6.24$ &  111.08  &  6.18    & 1\,$10^{-14}$   &
0.225   &   $9.22\,10^{-10}$\\
$\gamma$~Velorum    &$a=4.19, b=11.51, c= 12.2$&   50  &  2.58    & $3.4
\,10^{-4}$      & 0.101    & 0.014 \\
Berkeley~59 &$a=5.73  , b=19.57, c=14.46$  &  54.14  & 2.83     & $8.1\,10^{-5}$   
&  0.086    & $3.2\,10^{-3}$ \\
\hline
\end{tabular}
\end{center}
\end{table}
Table \ref {chi2lindleyweighted} for the \weighted{}. 
\begin{table}[ht!]
\caption
{
Numerical values of
$\chi_{red}^2$, AIC, 
probability $Q$,
$D$, 
the maximum distance between theoretical and observed CDF,
and  $P_{KS}$, 
significance level,   in the K--S test of the 
\weighted{} distribution  with  two parameters 
for  different  mass distributions.
The  number of  linear   bins, $n$, is 20.
}
\label{chi2lindleyweighted}
\begin{center}
\begin{tabular}{|c|c|c|c|c|c|c|}
\hline
Cluster     &parameters &  AIC    & $\chi_{red}^2$ & $Q$               &  $D$       &   $P_{KS}$      \\
\hline
NGC~2362    &b=0.008, c= 3.889  &   59.72  &  3.09   & 9.85\,$10^{-6}$ &
0.155    &   3.33\,$10^{-6}$ \\
NGC~6611    &b=1.57 , c=3.77 &  68.46  &  3.58    & 3.81\,$10^{-7}$   &
0.12    & 4.2 \,$10^{-3}$\\
$\gamma$~Velorum    &b= 0.0027, c= 5.86 & 79  &   4.16    & $6.2 \,10^{-9}$      
&  0.195    & $1.86\,10^{-8}$ \\
Berkeley~59 &b=0.007, c=  5.015  & 95.13  &  5.06     & $9\,10^{-12}$   
& 0.19     & $ 4.73\,10^{-15}$ \\
\hline
\end{tabular}
\end{center}
\end{table}
The best fit  for   NGC 2362
is obtained with  the \power{}, 
see  
Figure  \ref{lindley_power_ngc2362}.  

\begin{figure*}
\begin{center}
\includegraphics[width=6cm,angle=-90]{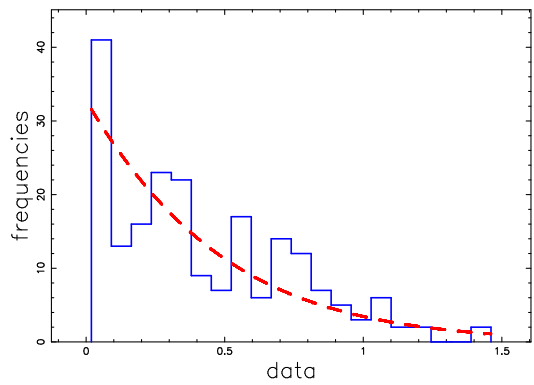}
\end{center}
\caption
{
Empirical PDF   of  mass distribution
for   NGC 6611 cluster data 
when the number of bins, $n$, is 20
(steps with blue full line)
with a superposition of the Lindley PDF with one parameter  
(red dashed line).
Theoretical parameters as in Table \ref{chi2lindleynorm}.
}
\label{lindley_ngc6611}
\end{figure*}

The best fit  for   NGC 6611
is obtained with  the Lindley PDF with one parameter, 
see  
Figure  \ref{lindley_ngc6611}.  

\begin{figure*}
\begin{center}
\includegraphics[width=6cm,angle=-90]{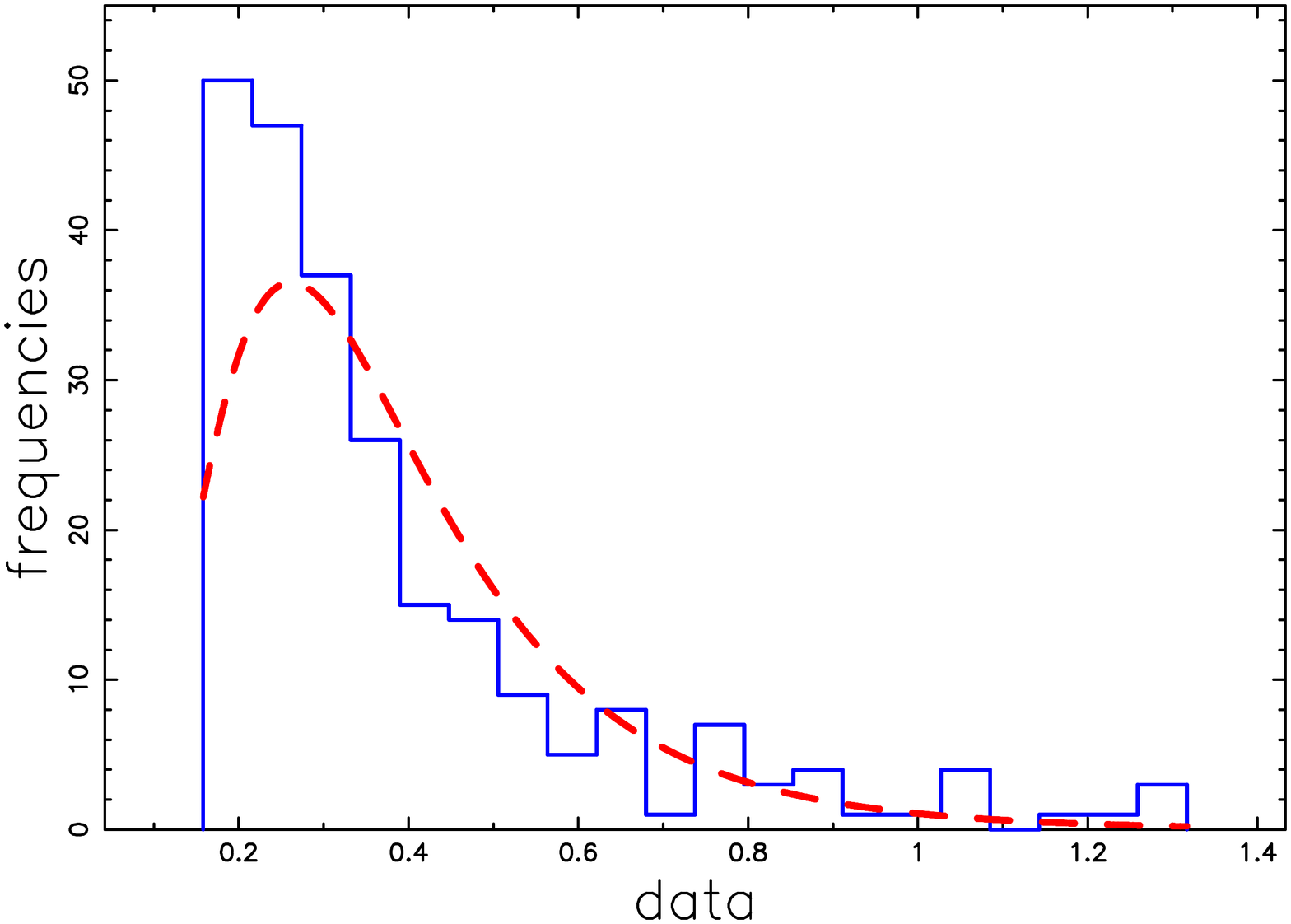}
\end{center}
\caption
{
Empirical PDF   of  mass distribution
for   $\gamma$ Velorum cluster data 
when the number of bins, $n$, is 20
(steps with blue full line)
with a superposition of the lognormal PDF  
(red dashed line).
Theoretical parameters as in Table \ref{chi2valueslognormal}.
}
\label{lognorm_gamma_vel}
\end{figure*}

The best fit  for  $\gamma$ Velorum
is obtained with  the lognormal  PDF, 
see  
Figure  \ref{lognorm_gamma_vel}.  

\begin{figure*}
\begin{center}
\includegraphics[width=6cm,angle=-90]{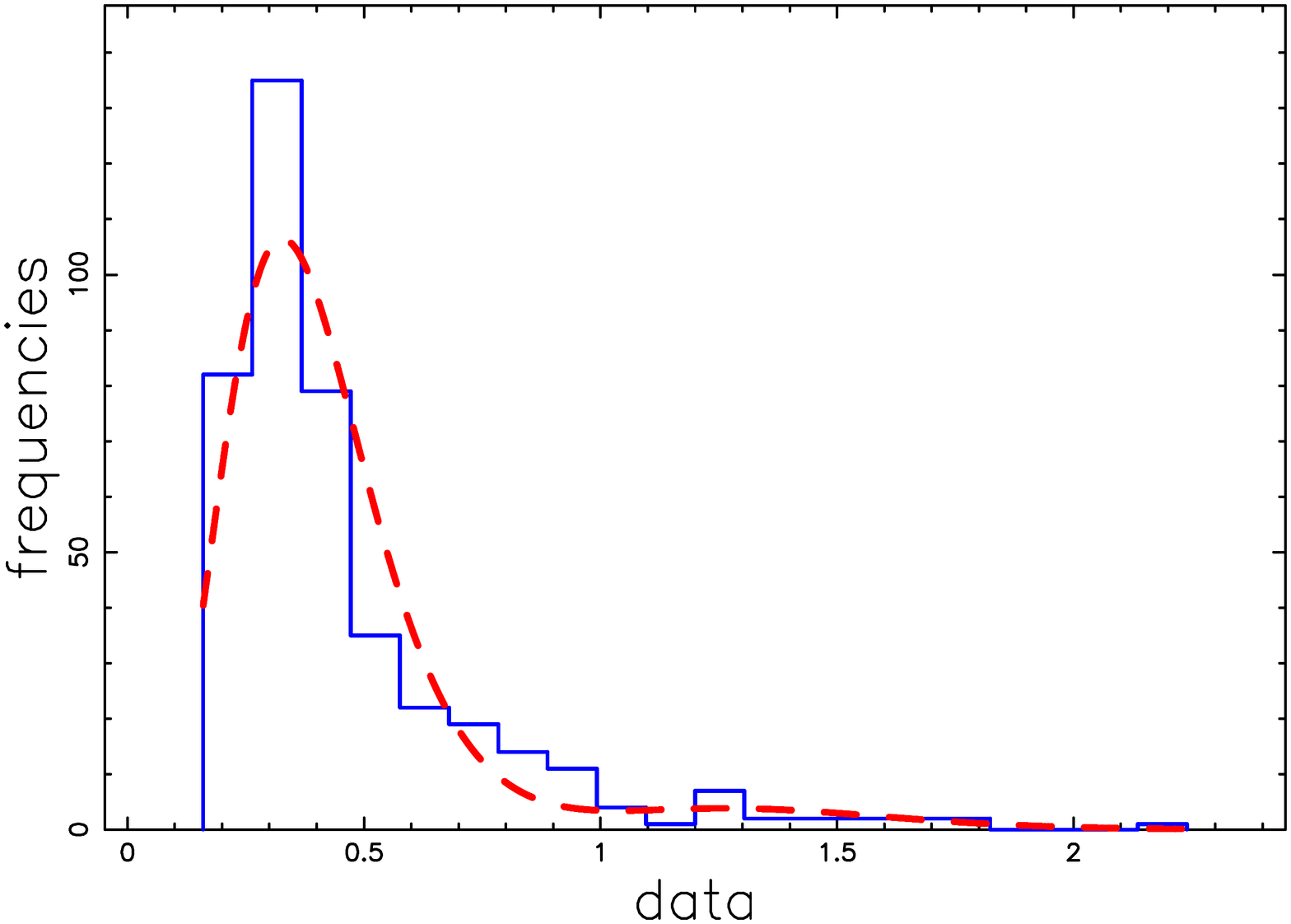}
\end{center}
\caption
{
Empirical PDF   of  mass distribution
for   the young cluster Berkeley 59 
when the number of bins, $n$, is 20
(steps with blue full line)
with a superposition of the \newgen{}  
(red dashed line).
Theoretical parameters as in Table \ref{chi2lindleynewgeneralized}.
}
\label{lindley_new_generalized_berkeley59}
\end{figure*}

The best fit  for the young cluster Berkeley 59
is obtained with  the \newgen{}, 
see  Figure  \ref{lindley_new_generalized_berkeley59}.  
\subsection{The double truncated Lindley}

\begin{figure*}
\begin{center}
\includegraphics[width=6cm,angle=-90]{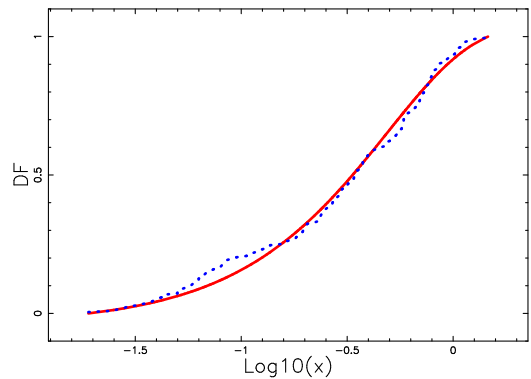}
\end{center}
\caption
{
Empirical CDF   of  mass distribution
for   NGC 6611 cluster data 
(blue dotted line)
with a superposition of the \truncated{} CDF with one parameter  
(red  line).
Theoretical parameters as in Table \ref{chi2lindleynormtrunc}.
}
\label{lindley_uno_trunc_df_ngc6611}
\end{figure*}

The  statistics for the \truncated{} 
with three  parameters are reported 
in Table \ref{chi2lindleynormtrunc}.
\begin{table}[ht!]
\caption
{
Numerical values of
$\chi_{red}^2$, AIC, 
probability $Q$,
$D$, 
the maximum distance between theoretical and observed CDF,
and  $P_{KS}$, 
significance level,   in the K--S test of the 
\truncated{} 
for  different  mass distributions.
The  number of  linear   bins, $n$, is 20.
}
\label{chi2lindleynormtrunc}
\begin{center}
\begin{tabular}{|c|c|c|c|c|c|c|}
\hline
Cluster     &parameters &  AIC    & $\chi_{red}^2$ & $Q$               &  $D$       &   $P_{KS}$      \\
\hline
NGC~2362    &$c= 1.61$, $x_l= 0.12$, $x_u=1.61$   &  156.7  &  8.86        & 
1.75\,$10^{-23}$  &  0.115    & 1.2\,$10^{-3}$\\
NGC~6611    &$c= 2.71$, $x_l= 0.019$, $x_u=1.46$    &  45.38  &  2.31          
&   0.0015           & 0.061    & 0.395 \\
$\gamma$~Velorum    &$c=  4.81$, $x_l= 0.158$, $x_u=1.317$   &   
45.89  &  2.34          & $1.3\,10^{-3}$  & 0.064    & 0.269 \\
Berkeley~59 &$c= 3.93$,  $x_l= 0.16$, $x_u= 2.24$   &   78.57  &  4.26
& $7.73\,10^{-9}$  & 0.134     & $4.35\,10^{-7}$ \\
\hline
\end{tabular}
\end{center}
\end{table}
Figure  \ref{lindley_uno_trunc_df_ngc6611}
reports the CDF of  the \truncated{}
for   NGC 6611
which is  the the best fit of the various distributions
here analysed for this cluster.   

\section{Conclusions}

\begin{figure*}
\begin{center}
\includegraphics[width=6cm,angle=-90]{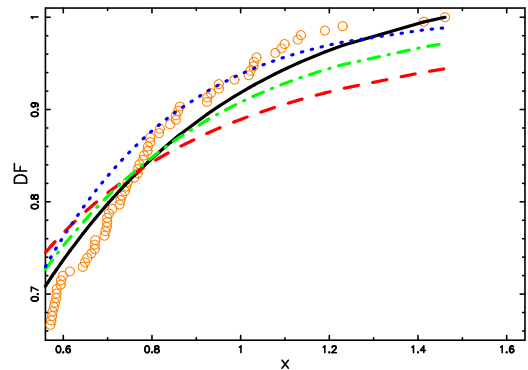}
\end{center}
\caption
{
Part of the empirical CDF   of  mass distribution
for   NGC 6611 cluster data 
(orange circles)
with a superposition of the \truncated{} CDF with one parameter  
(black full  line),
the  lognormal (red dashed line),
the  Lindley with one parameter (green dot-dash-dot-dash line)
and
the  \twopar{} (blue dot line).
}
\label{lindley_tutte}
\end{figure*}

In this  paper we explored five generalizations of
the Lindley distribution
as well the double truncated Lindley distribution
against the lognormal distribution.
For each  IMF of the four clusters here analysed, the distribution which realizes the best fit  
is reported in Table \ref{conclusions_fit}.
\begin{table}[ht!]
\caption
{
Best fits: Name of the cluster,
name of the distribution,
$D$, 
the maximum distance between theoretical and observed CDF,
and  $P_{KS}$, 
significance level,   in the K--S test.
}
\label{conclusions_fit}
\begin{center}
\begin{tabular}{|c|c|c|c|}
\hline
Cluster     & Best fit & $D$ & $P_{KS}$  \\
\hline
NGC~2362    &\power   & 0.053    &   0.39 \\
NGC~6611    & \truncated  & 0.061    & 0.395 \\
$\gamma$~Velorum   &\truncated &  0.064    & 0.269 \\
Berkeley~59        &\weighted  &   0.086    & $3.2\,10^{-3}$ \\
\hline
\end{tabular}
\end{center}
\end{table}
The above table allows to  conclude that 
the Lindley family here suggested produces better fits than
does the lognormal distribution.
Figure \ref{lindley_tutte} reports  
the CDF for NGC 6611 as well as four fitting 
curves.

\providecommand{\newblock}{}

\end{document}